\documentclass[conference]{IEEEtran}
\IEEEoverridecommandlockouts

\usepackage{cite}
\usepackage{amsmath,amssymb,amsfonts}
\usepackage{array} 
\usepackage{booktabs}
\usepackage{hyperref}
\usepackage{caption}  
\usepackage{algorithmic}
\usepackage{graphicx}
\usepackage{textcomp}
\usepackage{xcolor}

\def\BibTeX{{\rm B\kern-.05em{\sc i\kern-.025em b}\kern-.08em
    T\kern-.1667em\lower.7ex\hbox{E}\kern-.125emX}}
\begin{document}

\title{A Decision Support Framework for Blockchain Pattern Selection Based on Soft Goals
}

\author{
\IEEEauthorblockN{1\textsuperscript{st} Eddy Kiomba Kambilo}
\IEEEauthorblockA{\textit{Centre de Recherche en Informatique} \\
\textit{University Paris 1 Panthéon-Sorbonne} \\
Paris, France \\
eddy.kiomba-kambilo@univ-paris1.fr}
\and
\IEEEauthorblockN{2\textsuperscript{nd} Nicolas Herbaut}
\IEEEauthorblockA{\textit{Centre de Recherche en Informatique} \\
\textit{University Paris 1 Panthéon-Sorbonne} \\
Paris, France \\
nicolas.herbaut@univ-paris1.fr}
\and
\IEEEauthorblockN{3\textsuperscript{rd} Irina Rychkova}
\IEEEauthorblockA{\textit{Centre de Recherche en Informatique} \\
\textit{University Paris 1 Panthéon-Sorbonne} \\
Paris, France \\
irina.rychkova@univ-paris1.fr}
\and
\IEEEauthorblockN{4\textsuperscript{th} Carine Souveyet}
\textit{University Paris 1 Panthéon-Sorbonne} \\
Paris, France \\
carine.souveyet@univ-paris1.fr
}

\maketitle
\begin{abstract}
Blockchain technology is gaining momentum across many sectors. 
Whereas blockchain solutions have important positive effects on the business domain, 
they also introduce constraints and may cause delayed or unforeseen negative effects, 
undermining business strategies. The diversity of blockchain patterns and lack of standardized 
frameworks linking business goals to technical design decisions make pattern selection a 
complex task for system architects.

To address this challenge, we propose Blockchain--Technology-Aware Enterprise Modeling (BC-TEAEM), 
a decision support framework that combines ontologies of blockchain patterns 
and domain-independent soft goals with a multi-criteria decision-making 
approach. The framework focuses on the interplay between a domain expert 
and a technical expert to ensure alignment and traceability. By iteratively 
capturing and refining preferences, BC-TEAEM supports systematic selection 
of blockchain patterns.
We develop a prototype decision support tool implementing our method and 
validate it through a case study of a pharmaceutical company’s supply chain 
traceability system, demonstrating the framework’s applicability.

\end{abstract}

\begin{IEEEkeywords}
Blockchain patterns, enterprise architecture, soft goals, decision-support
\end{IEEEkeywords}

\section{Introduction}
Blockchain technology is increasingly being adopted in inter-organizational and distributed systems 
to provide trust, transparency, traceability, and immutability. This adoption has led to a 
growing body of reusable blockchain patterns, each offering distinct advantages and trade-offs 
with respect to key soft goals (non-functional requirements)—such as performance, security, scalability, and cost.
However, selecting an appropriate set of patterns remains a non-trivial challenge.
Patterns often have overlapping or conflicting effects, and their selection is typically driven by 
the architects' experience rather than a systematic process. This can result in misalignment between 
business goals and technical decisions, producing costly or suboptimal designs.

\vspace{0.1cm}
Existing works on blockchain patterns have focused on compiling descriptive catalogs~\cite{xu2018pattern},
or building pattern ontologies~\cite{six2022ontology}.
While valuable, these approaches rarely link patterns to soft goals in a way that supports automated reasoning 
or dynamic trade-off analysis. From a requirements engineering perspective, methods for soft goal modeling exist,
 but their integration into blockchain-specific decision support systems remains underexplored.
Multi-criteria decision-making (MCDM) methods have been applied in related contexts\cite{summa2023}, 
yet they typically rely on static attributes or rule-based reasoning, offering limited adaptability to changing 
preferences or complex interdependencies.

\vspace{0.1cm} 
To illustrate the challenge, consider a pharmaceutical company that aims to enhance trust in its supply chain processes 
through blockchain-based drug traceability. The business expert prioritizes security (ensuring shipment record integrity), 
performance ($\geq$ 1,000 TPS to match legacy systems), and cost reduction. 
Meeting these goals requires patterns that address functional needs while also balancing trade-offs between conflicting
soft goals—for instance, an encryption pattern may strengthen security but reduce performance. Without structured decision support, 
such trade-offs are handled ad hoc, increasing the risk of inconsistent or suboptimal outcomes.

\vspace{0.1cm} 
In this work, we investigate how to systematically connect business-level soft goals to blockchain design patterns in order
to improve the transparency, traceability, and justifiability of architectural choices.
We address three research questions(RQs):
\begin{itemize}
    \item \textbf{RQ1:} How can soft goals defined at the business level be explicitly aligned with
    blockchain patterns in the technical domain?
    \item \textbf{RQ2:} How can the impact of specific technical decisions (e.g., choosing a blockchain pattern) be evaluated 
    and traced back to business goals?
    \item \textbf{RQ3:} How should domain experts and technical experts interact to ensure a
    business goal–driven and technology-aware solution design?
\end{itemize}

To explore these questions, we propose BC-TEAEM, a decision-support framework that combines a domain-specific ontology,
soft goal modeling, and an MCDM method for pattern ranking. 
The framework supports iterative refinement of preferences, explicit trade-off reasoning, and traceable recommendations. 
We validate our approach with a supply-chain traceability case study and mixed-methods evaluation.

\vspace{0.1cm} 
By following Wieringa's perspective on DSR~\cite{wieringa2014design}, we ensure both practical relevance and scientific rigor. 
Our approach follows stages of the DSR cycle, from problem identification to Evaluation. 
The remainder of this paper is organized as follows.

\begin{itemize}
    \item \textbf{Problem Investigation}: Section~\ref{probleminvestigation} reviews existing literature and analyzes current 
    limitations in selecting blockchain patterns based on soft goals.
    In Section~\ref{artifactdesignmethology}, we present our artifact design methodology addressing these limitations.
    
    \vspace{0.1cm} \noindent
    \item \textbf{Artifact Conception and Development}: In section \ref{designDev}, we present the proposed artifact, 
    which models the relationships between blockchain patterns and soft goals to support architectural decisions.
    We also describe the development of an interactive recommendation tool that integrates this ontology with a 
    Multi-Criteria Decision-Making (MCDM) method.

    \vspace{0.1cm}  
    \item \textbf{Evaluation/Validation}: Section~\ref{evaluationprocess} illustrates the application of the tool through 
    scenario-based evaluations. 
    We assess its relevance, usability, and recommendation quality based on scenario-based experimentation and expert feedback. 
    Section~\ref{discussion} provides a discussion of the findings, and Section~\ref{conclusion} concludes the paper.
\end{itemize}

\section{Problem Investigation \& Research Gap}\label{probleminvestigation}
Unlike functional requirements (hard goals), soft goals are qualitative, subjective and context-dependent.
They can be satisfied to varying degrees and often involve trade-offs~\cite{hu2015semantic}. 
Selecting blockchain patterns based on soft goals is challenging because a single pattern can influence multiple goals in different,
sometimes conflicting, ways. 
For example, Encrypting On-chain Data improves confidentiality but may introduce latency, 
thereby degrading performance, while Off-Chain Data Storage can improve performance but may weaken confidentiality due 
to reliance on external systems.

\vspace{0.1cm}
A number of works have documented blockchain patterns. Extensive catalogs 
organized into categories such as interaction, data management, security, 
and structural patterns are presented in \cite{xu2018pattern}, \cite{xu2025pattern}.
Building on this, a blockchain pattern ontology is proposed to formalize structural 
relationships between patterns~\cite{six2022ontology}. Another ontology focuses
on decentralized application (dApp) development, emphasizing functional 
components rather than non-functional trade-offs~\cite{Besancon2022}.
While these resources are valuable, they primarily serve as descriptive references
and do not connect patterns to soft goals in a way that supports automated 
reasoning.


\vspace{0.1cm}
From a requirements engineering perspective, soft goal modeling has been widely 
studied~\cite{hu2015semantic},\cite{sumesh2022}, often using frameworks such as
the NFR (Non-Functional Requirements) Framework to capture qualitative attributes
and interdependencies.
However, few works explicitly integrate these models into blockchain-specific 
architectural decision-making. The importance of managing trade-offs in blockchain
ecosystems is highlighted by~\cite{janovic}, but without providing a formalized
link between soft goals and patterns.

\vspace{0.1cm}
Decision-support tools for blockchain-related design have also emerged.
Multi-Criteria Decision-Making (MCDM) techniques are applied to platform or 
configuration selection~\cite{bahar2023dynamic},\cite{summa2023},while 
TOPSIS is explored for blockchain sharding strategies~\cite{Lui2022}.
These approaches demonstrate the utility of quantitative trade-off analysis but
typically rely on static attributes, lack integration with ontologies, and do 
not model inter-pattern or inter-soft goal relationships.

Overall, existing approaches tend to address either:
\begin{enumerate}
    \item \textbf{Pattern representation} — focusing on cataloging or structural 
    ontologies without modeling impacts on soft goals, or
    \item \textbf{Soft goal modeling} — emphasizing qualitative attributes and 
    interdependencies without linking to specific patterns.
    \item \textbf{Decision-making} — applying MCDM methods without leveraging an 
    ontology to capture complex interdependencies.
\end{enumerate}

In contrast to prior work, we propose BC-TEAEM, which fills this gap by combining 
three key elements: 
(i) formalizes the relationships between blockchain patterns and soft goals 
through a domain-specific ontology, building upon foundational principles from 
the ontology engineering work of~\cite{guizzardi2015towards}
(ii) a user interface for capturing and processing soft goal preferences, 
and (iii) a recommendation engine using MCDM to support trade-off analysis. 

\section{Artifact Design Methodology}\label{artifactdesignmethology}

\subsection{Selection of Blockchain Patterns and Soft Goals}
Blockchain patterns and soft goals are concepts that may be defined differently 
across domains. In this work, we focus on blockchain-based applications and 
provide contextualized definitions of each selected soft goal.

\vspace{0.1cm}
To identify relevant patterns, we began with catalogs proposed by 
~\cite{xu2018pattern},\cite{xu2025pattern}, which provides a list of 15 
blockchain patterns divided into four categories. 
\textit{Interaction between blockchain and external world}, which describes 
blockchain communication with external systems.
\textit{Data management patterns}, which address data storage (on-chain/off-chain storage). 
\textit{Security patterns} that focus on the security aspects of blockchain-based applications; 
and \textit{Structural patterns}, which describe smart contract dependencies 
and behaviors.
We extended this set with variants of patterns reported in more recent studies 
\cite{rajasek},\cite{henry2024} and industrial practice 
(e.g., centralized oracle, zero-knowledge proof–based privacy).
A subset of these patterns is presented in Table~\ref{tab:relationships};
The complete list and descriptions are provided in \cite{xu2018pattern}.

\vspace{0.1cm}
The soft goals considered in our framework draw on established requirements 
engineering frameworks such as the NFR Framework\cite{hu2015semantic} and 
ISO/IEC 25010 quality model, as well as blockchain-specific literature~\cite{kambilo2023}. 
We selected goals frequently cited in both research and practice for blockchain 
system design: 
\textit{Cost}: related to transaction fees and infrastructure costs 
(deployment and maintenance).
\textit{Integrity}: the assurance of data consistency, immutability, and 
verifiability across the blockchain network.
\textit{Performance}: the ability to process transactions efficiently at scale 
(e.g., transaction speed, throughput, latency, congestion).
\textit{Interoperability}:the system’s capacity to communicate with external 
systems (e.g., IoT devices, databases).
\textit{Privacy}: the ability to control access to sensitive data while 
maintaining transparency where necessary. 
\textit{Security}: protection against cyberattacks, cryptographic threats, 
unauthorized access, and other malicious activities.
\textit{Transparency}: the provision of public visibility and auditability of 
data for all stakeholders.

This choice ensures coverage of key non-functional concerns relevant to 
trustworthiness in blockchain-enabled systems.

\subsection{Modeling Trade-offs Between Patterns and Soft Goals}\label{tradeoffs}
Patterns may be mutually supportive or conflicting, making it essential to 
capture both inter-pattern and inter–soft goal relationships.
Building on~\cite{six2022}, we extend existing relationship types with:
 
 \begin{itemize}
    \item \textbf{Benefits from:} one pattern may reinforce or complement another. 
    For example, the Decentralized Oracle pattern can benefit the Off-Chain Data 
    Storage pattern by providing a secure and reliable way to fetch external 
    data, thereby enhancing the system’s interoperability and performance.

    \item \textbf{Conflicts with}: certain patterns may constrain the use of 
    others. Their combination can lead to contradictory systems or design 
    inconsistencies. This type of relationship is useful for identifying 
    potential design trade-offs, which are important in pattern recommendation.
    For example, Encrypting On-Chain Data improves privacy, while Reverse Oracle 
    reduces it by exposing data externally.
\end{itemize} 

\subsection{Modeling the Relationship Between Patterns and soft goals}\label{express}
To capture the relationship between patterns and soft goals, we conducted an 
engineering analysis grounded in the foundational works \cite{xu2018pattern}\cite{xu2025pattern}. 
Their studies provide valuable contextual insights and taxonomic classifications, 
but they do not include formal representation suitable for automated reasoning.

\begin{table}[ht]
\centering
\caption{Scoring Process: Influence of Patterns on Soft Goals}
\label{tab:scoring_process}
\begin{tabular}{p{0.8cm}p{1.5cm}p{4.7cm}}
\hline
\textbf{Symbol} & \textbf{Influence Type} & \textbf{Definition} \\
\hline
++ & Strong \newline Positive & Pattern significantly improves the soft goal under typical usage scenarios. \\
\hline
+ & Moderate Positive & Pattern improves the soft goal, but the gains are limited or context-dependent. \\
\hline
0 & Neutral / No Measurable & Pattern has a negligible or no measurable impact on the soft goal. \\
\hline
- & Moderate Negatives & Pattern somewhat hinders the soft goal, with limited severity. \\
\hline
--- & Strong Negative & Pattern significantly worsens the soft goal.\\
\hline
\end{tabular}
\end{table}
 
Each influence is rated on a 5-point ordinal scale, with the criteria shown in 
Table~\ref{tab:scoring_process}:

The scoring process is designed to capture both positive and negative influences 
of patterns on soft goals, The scoring process is conducted as follows::
\begin{itemize}
    \item Review literature descriptions (e.g.~\cite{xu2025pattern}~\cite{xu2018pattern}) 
    for each pattern’s forces and consequences.
    \item Identify direct and indirect effects on each soft goal.
    \item Assign scores using Table~\ref{tab:scoring_process} criteria. 
    \item Where evidence is ambiguous, validate scores through expert 
    consultation or workshop discussions.
\end{itemize}

This explicit methodology reduces subjectivity and allows future researchers 
to reproduce or refine the scoring process; and is one of the key contributions
of our work.

\noindent This systematic encoding produces a formal knowledge base of 
pattern-soft goal relationships. 
The resulting model is specifically engineered for MCDM integration, enabling 
reasoning and trade-offs analysis in pattern selection.
\begin{table}[htbp]
\centering
\caption{Relationships Between Patterns and soft goals}
\label{tab:relationships}
\begin{tabular}{
    >{\raggedright\arraybackslash}p{1.6cm}
    *{7}{c}
}
\toprule
\textbf{Patterns} & \multicolumn{7}{c}{\textbf{Soft Goals}} \\
\cmidrule(lr){2-8}
 & \textbf{Cost} & \textbf{Inter.} & \textbf{Priv.} & \textbf{Secu.} & \textbf{Integ.} & \textbf{Trans.} & \textbf{Perf.} \\
\midrule
\scriptsize Oracle & N & I & W & W & N & W & W \\ 
\scriptsize Reverse Oracle & N & I & N & W & N & N & I \\
\scriptsize Encrypting On-Chain & N & N & I & W & N & N & W \\
\scriptsize Off-chain Data Storage & I & N & N & W & I & W & I \\
\rotatebox[origin=c]{90}{\dots}  & \rotatebox[origin=c]{90}{\dots}  & \rotatebox[origin=c]{90}{\dots}  & \rotatebox[origin=c]{90}{\dots}  & \rotatebox[origin=c]{90}{\dots}  & \rotatebox[origin=c]{90}{\dots}  & \rotatebox[origin=c]{90}{\dots}  & \rotatebox[origin=c]{90}{\dots} \\

Factory contract & W & N & N & I & N & N & I \\
\bottomrule
\end{tabular}

\vspace{0.1cm}
\footnotesize
\textbf{Legend:} 
I = Strong and Moderately Improves;
W = Strong  and Moderately Worsens; N = Neutral.  
\\
The abbreviations: Inter. (Interoperability), Trans. (Transparency), Perf. (Performance).
Priv. (Privacy), Secu. (Security), Integ. (Integrity).
\end{table}

Table \ref{tab:relationships} summarizes the encoded relationships between selected patterns and soft goals. 
For example, Oracle patterns strongly improve interoperability but moderately worsen security due to external data dependencies. 
In contrast, Encrypting On-Chain Data strongly improves privacy but moderately worsens performance. Notably, 
some patterns (e.g., Factory Contract) have largely neutral effects on most soft goals, suggesting they may 
be selected with minimal trade-offs.
These observations highlight both expected and counter-intuitive relationships—e.g., Off-Chain Data Storage improves performance 
but can weaken integrity, a trade-off often overlooked in practice. Such insights motivate the need for systematic reasoning 
in pattern selection.
Detailed information on how each pattern impacts the soft goals is
available in the supplementary document\footnote{\url{https://github.com/CRI-Collab/ontologyDesccription}}

\begin{figure*}
    \centering
    \includegraphics[scale=0.43]{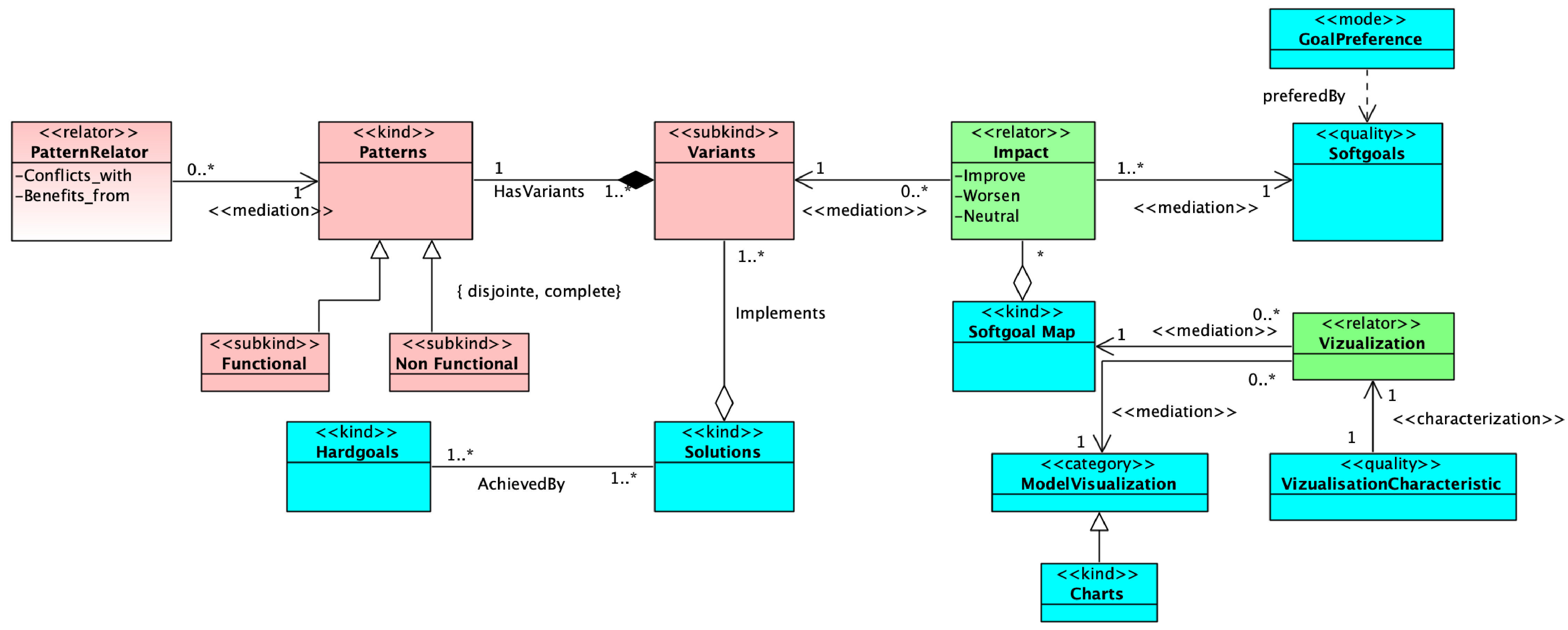}
    \caption{This figure illustrates the ontology of the relationship between the blockchain patterns and the soft goals.}
    \label{fig:ontology}
\end{figure*}

\subsection{MCDM Selection}\label{topsis}
To support trade-off resolution and pattern ranking, we opt for the Technique 
for Order of Preference by Similarity to Ideal Solution (TOPSIS)~\cite{papathanasiou2018topsis} 
as the Multi-Criteria Decision-Making method due to its simplicity, robustness, 
and relevance in goal-oriented trade-off analysis. 

Unlike qualitative ranking methods, TOPSIS quantitatively evaluates each 
pattern by calculating its geometric distance from both an ideal and an 
anti-ideal solution, based on user-defined soft goal weights. This makes it 
particularly suitable for scenarios involving conflicting non-functional 
requirements.

TOPSIS is especially adapted in our context for the following reasons: 
(a) It supports dynamic user input and real-time re-ranking of patterns as 
stakeholder preferences evolve. 
(b) It aligns with the concept of soft goal trade-offs by capturing closeness 
to an ideal solution. 
(c) It also produces a continuous score that facilitates the ranking and 
visualization of multiple pattern alternatives.

The scoring matrix from Section~\ref{express} serves as the decision matrix 
for TOPSIS, with soft goal weights set according to domain expert soft goal 
priorities.

\section{Artifact conception and Development}\label{designDev}
BC-TEAEM consists of two interrelated artifacts: (1) a domain-specific ontology 
modeling the relationships between blockchain patterns and soft goals, and 
(2) an interactive tool that operationalizes this ontology for pattern 
recommendation.

\subsection{Ontology Construction}
The ontology is designed using OntoUML~\cite{guizzardi2015towards} and builds 
on the blockchain software patterns ontology by~\cite{six2022ontology}.
This ontology defines relationships such as: ``Created from'', ``Variant of'', 
``Requires'', and ``Related to''. We extend it by:
\begin{itemize}
    \item Adding impact relationships between patterns (or their variants) and soft goals, 
    with direction and magnitude encoded using the scoring criteria from 
    Section \ref{express}.
    \item Introducing inter-pattern relationships: ``Conflict With'' and 
    ``Benefits From'' defined in Section \ref{tradeoffs}.
\end{itemize}

\textbf{\textit{To answer RQ$_1$}}, the first outcome is the creation of the 
TEAEM ontology, which formalizes the relationships between soft goals 
(business goals) and blockchain patterns, and models trade-offs among patterns
(top-down alignment).
As shown in Figure~\ref{fig:ontology}, the driving idea behind the ontology is: 
(a) The base structure (highlighted in pink) is derived from the ontology
proposed by~\cite{six2022ontology}. It includes foundational concepts such as 
Pattern, Variants, and their general relationships.
(b) Our Contribution extends this foundation through additional components 
(blue and green) that model trade-offs between patterns and integrate soft 
goals and their influence.
At the core of the ontology is the \textit{\textbf{Impact}} Relator, which 
links pattern Variants to soft goals, and their types of influence.

\vspace{0.1cm}
\textit{\textbf{Patterns Class}} represents blockchain patterns 
(Table~\ref{tab:relationships}), organized into two categories: 
\begin{itemize} 
    \item \textit{Functional patterns:} Core capabilities (for example, Oracle patterns for external data).
    \item \textit{Non-functional patterns:} Quality attributes (for example, Multiple Authorization for security). 
\end{itemize} 

\vspace{0.1cm}
In our running example, architects use the Oracle pattern to integrate 
off-chain data. Two variants are possible:
\begin{itemize}
    \item Centralized oracle — Improves performance and reduces cost, 
    but reduces security due to a single point of failure.
    \item Decentralized oracle — Improves security but worsens performance 
    due to additional validation steps.
\end{itemize}

The ontology can detect such conflicts automatically. 
For example, if two variants of the same pattern have opposing influences on 
the same soft goal and $v1 \neq v2$, the ontology asserts a conflict relation 
between them.

\vspace{0.1cm}
\textit{\textbf{Softgoals class}} represents qualitative attributes such as 
performance and security (defined in Section~\ref{artifactdesignmethology}). 
Each \textit{\textbf{Variant}} is connected to \textit{soft goals} through the
\textit{\textbf{Impact}} Relator. These links are interpreted in light of user
preferences.
\noindent In our running example, the supply chain enterprise gives higher 
priority to \textit{Security} than to \textit{Performance}.
The decentralized oracle improves security but worsens performance. 
According to the ontology's reasoning process, the trade-offs introduced by 
the decentralized oracle variant are considered acceptable. 
This implies that the benefits of enhanced security outweigh the drawbacks of
reduced performance. Therefore, this variant is recommended.

\vspace{0.1cm}
For Reasoning on Pattern Interactions, in addition to pattern-soft goal 
relationships, the ontology models inter-pattern relationships. 
Some patterns have synergistic impacts; others are conflicting. For example,
Encrypting On-chain Data improves privacy but degrades performance, 
and Off-chain Data storage improves performance but may weaken data transparency. 
Their combined selection creates a conflict effect, which the ontology can 
detect automatically. The reasoning process helps blockchain architects spot 
potential conflicts early to prevent unexpected issues in the final design of 
the system.

\vspace{0.1cm}
The outcome of the reasoning process is a set of recommended solutions. 
These are captured in the \textit{\textbf{Solutions}} class, which encapsulates the concrete implementation of patterns and their variants. 
All impacts are aggregated and structured within the \textit{\textbf{Softgoal Map}} class, instantiated in the ontology, and linked 
to higher-level goals represented by the \textit{\textbf{Hardgoals}} class.%

We assessed the ontology’s correctness and completeness using a standard technique:
\begin{itemize}
    \item Consistency Checking — Performed in Protégé using HermiT reasoner; 
    no unsatisfiable classes or inconsistent individuals detected.
\end{itemize}

\subsection{BC-TEAEM Tool (artifact 2)}
In parallel with the ontology, we develop a tool that provides a user-friendly interface to interact with the ontology,
 avoiding the need for domain experts to use ontology editors such as Protégé.

\begin{figure*}
\centerline{\includegraphics[scale=0.40]{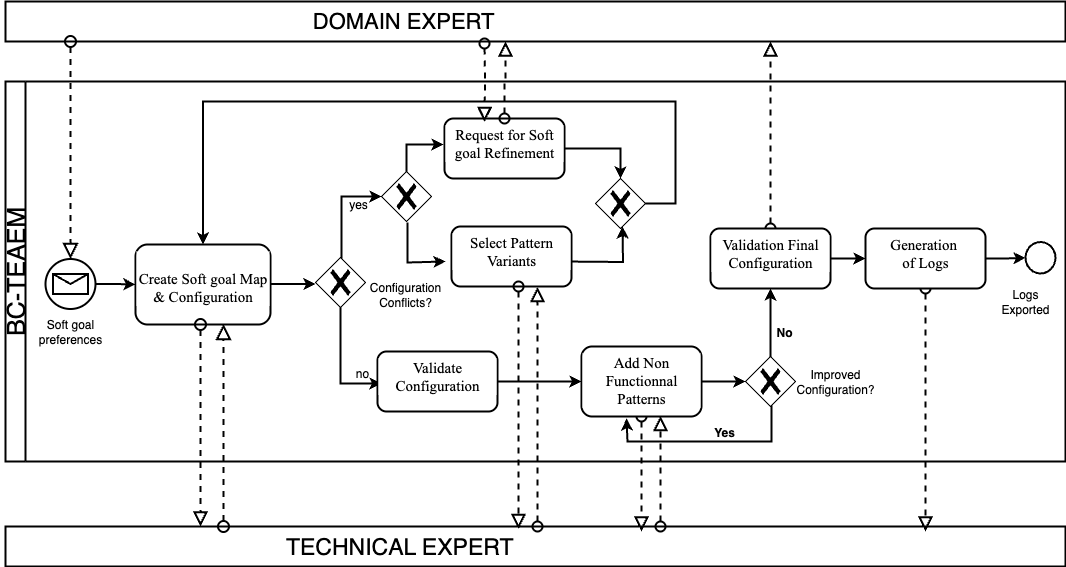}}
\caption{This figure illustrates the workflow of the applications by considering users' roles and tools features.}\label{fig:process}
\end{figure*}

\vspace{0.1cm}
\textit{\textbf{To address RQ$_3$}}, we apply the workflow illustrated in Figure~\ref{fig:process} to our running example 
using the interactive pattern selection process.

\subsubsection{\textbf{User Roles}}
We distinguish two user profiles:
\begin{itemize}
    \item \textbf{Domain Experts (DE)}: focus on soft goals and trade-offs, providing high-level requirements and preferences.
    \item \textbf{Technical Experts (TE)}: refine, validate, and implement feasible pattern-based solutions based on DE input.
\end{itemize}

\subsubsection{\textbf{Tool Workflow}}
The tool follows the workflow described in Figure \ref{fig:process} consisting of:

\begin{itemize}
    \item[$\bullet$] \textit{\textbf{Explorer Module (Domain Experts)}}: Allows DE and TE to browse patterns, 
    view their soft goal impacts, and adjust preferences. 
    DE can specify their preferences in two ways
    
    \begin{itemize}
        \item \textit{Improves:} means that the domain expert is interested in patterns that positively contribute to a specific soft goal.
                      (weights=$+2$ (for strongly agree) and $+1$ (for agree)) 
        
        \item  \textit{Best effort:} Attempt to improve a soft goal; if not feasible, consider alternatives
        (weight adjusted automatically from $+2$ or $+1$ to $0$, $-1$, or $-2$.)
    \end{itemize}
\end{itemize}

\begin{itemize}
    \item[$\bullet$] \textit{\textbf{Recommendation Engine:}} This module processes soft goal preferences and constructs a visual representation 
    of different possible solutions as referred to as the \textit{Softgoal Map (SMp)} which highlights the alignment between patterns and prioritized soft goals.
     Based on the SMp, three main recommendation cases are supported:
    \begin{itemize}
        \item \textit{Single Match:} One Optimal solution in the SMp, the tool recommends it.
        \vspace{0.2cm}
        \item \textit{Multiple Matches:} If multiple patterns are identified, TE refines the recommendation using domain knowledge. 
         \vspace{0.2cm}
        \item \textit{Conflicting Goals:} If trade-offs exist (example, security vs performance),
         DE is consulted to adjust priorities(soft goals), triggering a recalculation (RQ$_2$).
    \end{itemize}
\end{itemize}

\subsubsection{\textbf{Resolving Trade-offs and Refining Solutions}}
To address trade-offs among competing soft goals, our approach relies on an
iterative refinement process. When the relative importance of soft goals  
(e.g., performance vs security) can not be determined by the technical expert, 
the domain expert is consulted to provide guidance (prioritization of soft goals).
Based on this input, the tool recalculates the soft goal satisfaction map 
and updates the recommended configuration.

\vspace{0.1cm}
This iterative process continues until a solution is found that balances 
functional and non-functional requirements. In doing so, the process 
directly addresses, \textit{\textbf{RQ$_2$} by enabling bottom-up traceability}:
each configuration refinement evaluates how technical decisions 
(e.g., pattern selection) impact business-level goals through the knowledge base.

\vspace{0.1cm}
The initial configuration includes only the functional patterns essential 
for implementing the application logic. In a second step, non-functional 
patterns are integrated to improve the overall solution in accordance with 
the prioritized soft goals. These patterns are evaluated and ranked using 
the TOPSIS algorithm, which computes a score based on their contribution 
to the selected goals.

\vspace{0.1cm}
To support decision-making, each non-functional pattern is presented in the interface as a visual card 
summarizing its impact on individual soft goals. 
Once a satisfactory configuration is reached, the tool may suggest additional, complementary patterns to further enhance the solution. 
This ensures that both functional and non-functional requirements are addressed in a balanced and goal-oriented manner. 
The tool is available online\footnote{\url{https://demo.ontoteaem.fr/}}, and the source code is accessible via GitHub\footnote{\url{https://github.com/CRI-Collab/ontoTEAEMS}}.

\section{Evaluation}\label{evaluationprocess}
\subsection{Evaluation Process}
Following design science principles, we evaluated BC-TEAEM using a mixed-methods approach that combines:
\begin{itemize}
    \item Scenario-based experimentation with domain-relevant tasks to assess recommendation relevance and trade-off clarity.
    \item Structured questionnaires to measure perceived usability, usefulness, and user confidence.
\end{itemize}
This dual method was chosen to capture both qualitative insights (e.g., expert reasoning, observed interactions)
and quantitative perceptions (e.g., Likert-Scale responses).

\noindent
To guide our evaluation, we formulated the following hypotheses:
\begin{itemize}
    \item[\textbf{H$_1$:}] \textit{The tool supports decision-making by providing relevant recommendations and clear trade-offs when used in realistic design scenarios.}  
    \item[\textbf{H$_2$:}] \textit{The tool is perceived as usable, useful, and understandable by users, including those with limited experience in blockchain pattern selection.}
\end{itemize}

\subsection{Scenario-based experimentation}
\noindent\textbf{Purpose:} Evaluate whether BC-TEAEM produces relevant 
recommendations and clearly communicates trade-offs in realistic design 
scenarios (addresses RQ$_1$ and RQ$_2$ and H$_1$).

\vspace{0.1cm}
\noindent\textbf{Data Collection:} The evaluation involved seven 
researchers with blockchain and/or software design experience (Table~\ref{tab:panel}).

\begin{table}[ht]
\centering
\caption{Participant Panel overview}
\begin{tabular}{lccccccc}
\toprule
\textbf{} & \textbf{E1} & \textbf{E2} & \textbf{E3} & \textbf{E4} & \textbf{E5} & \textbf{E6} & \textbf{E7} \\
\midrule
\textbf{Role} & R1 & R2 & R3 & R4 & R5 & R6 & R7 \\
\textbf{Blockchain Exp (y)} & 4 & 4 & 4 & 4 & 2 & 1 & 2 \\
\textbf{Software Design Exp} & 5+ & 1 & 5 & 5 & 2 & 2 & 5+ \\
\bottomrule
\end{tabular}
\label{tab:panel}
\end{table}

\vspace{0.1cm}
\noindent\textbf{Experimental Settings:} Each participant was presented 
with a scenario involving a logistics enterprise seeking to prioritize three 
soft goals: \textit{cost reduction, performance improvement, and security enhancement}. 
They used the tool to generate pattern recommendations and evaluate 
trade-offs between variants (e.g., Oracle vs. Reverse Oracle).
Assessing the tool’s output for relevance, clarity of trade-offs, and 
interface usability.

\vspace{0.1cm}
\noindent\textbf{Data Analysis \& Results:} The data collected from the 
semi-structured interviews and tool interactions (logs).
The results of the scenario-based evaluation showed that the tool 
successfully recommended patterns aligned with the specified soft goals.
The tool identified \textbf{Off-Chain Data Storage} as the most suitable 
pattern, and highlighted trade-offs between Oracle and Reverse Oracle 
pattern variants.
Responses were categorized into:
\begin{itemize}
    \item[$\bullet$] \textbf{Relevance of Recommendations:} Five of seven participants agreed aligned with industry practices (e.g., Off-Chain Data Storage). 
    Minor disagreements (E2, E6) highlighted contextual nuances in cost-performance trade-offs.
    \item[$\bullet$] \textbf{Clarity of Trade-offs:} 6/7 found them ``clearly presented and comprehensive''. E1 noted that additional examples could further improve clarity.
    \item[$\bullet$] \textbf{Usability of the Interface:} 4 experts found the interface intuitive for a 30-minute session, while 3 suggested streamlining terminology (e.g., "Reverse Oracle" required glossary support).
\end{itemize}

\noindent\textbf{Threats to Validity.}
\begin{itemize}
    \item[$\bullet$] \textbf{External:} Limited to supply chain scenarios;
    generalization requires broader domains.
    \item[$\bullet$] \textbf{Construct:} Time constraints (30 minutes) may
    not reflect real-world deliberation.
\end{itemize}

\subsection{Structured Questionnaire}
\noindent\textbf{Purpose:} To evaluate \textbf{H$_2$} and textbf{RQ$_3$}—the 
perceived usability, usefulness, and understandability of the tool by a 
broader audience, including less experienced users.

\vspace{0.1cm}
\noindent\textbf{Data Collection:} we collected data from 32 participants (26 postgraduate students, 4 researchers, and 2 industry experts). 
The Structured Google Forms questionnaire was based on TAM~\cite{tam2015} and SUS~\cite{sus2024}, assessing: 
\begin{itemize}
    \item[$\bullet$] \textit{Usability} which evaluates how easy and 
    intuitive the tool is to use. 
    \item[$\bullet$] \textit{Self-efficacy} measuring users' confidence in
    independently using the tool to achieve their goals (for instance, 
    exploring blockchain patterns), and 
    \item[$\bullet$] \textit{Usefulness of Features} assessing whether the 
    tool’s functionalities (recommendations, pattern comparisons) meet 
    users’ needs.
\end{itemize}

\vspace{0.1cm}
\noindent \textbf{Experimental Protocol:}
Participants interacted with the BC-TEAEM tool during a supervised lab session
exploring pattern recommendations for a given set of soft goals.
Participants rated the tool's usability, self-efficacy, and usefulness of features.

\vspace{0.1cm}
\noindent \textbf{Data Analysis \& Results:} Descriptive statistics were 
computed from responses (Table~\ref{tab:analysis}). This analysis helped us 
understand the overall perception of the tool and identify areas for 
improvement.

\begin{table}[h]
\centering
\caption{Perception Scores from Participants}\label{tab:analysis}
\begin{tabular}{p{3cm} p{2cm} p{2.6cm}}
\hline
\textbf{Category} & \textbf{Average/5} & \textbf{Standard Deviation} \\
\hline
\textbf{Usability } & 3.9  & 1.1 \\
\textbf{Self-Efficacy} & 3.4  & 1.2 \\
\textbf{Usefulness of Features} & 4.1  & 0.8 \\
\hline
\end{tabular}
\end{table}

\noindent The results indicate that the tool is perceived as highly useful 
(average of 4.1/5), with its blockchain pattern recommendations considered 
valuable for application design.
However, usability (3.9/5) and self-efficacy (3.4/5) scores reveal challenges 
in clarity and user confidence. 
The higher standard deviations for these categories reflect divergent user 
experiences—some found the tool intuitive, while others highlighted the need 
for additional guidance (e.g., video tutorials, tooltips) or training. Feedback
suggests prioritizing improvements such as clearer feature explanations and 
a more intuitive interface. Implementing these adjustments could enhance 
adoption by reducing the learning curve.

\vspace{0.1cm}
\noindent \textbf{Threats to Validity:}
\begin{itemize}
    \item[$\bullet$] \textbf{Sampling bias:} The sample over-represents 
    students (87\%), which may limit direct generalization to industry practice. 
    However, their diversity in technical backgrounds provided varied 
    perspectives on usability and clarity.
\end{itemize}

Preliminary results support H$_1$ and H$_2$, formal statistical testing
(t-tests against neutral baselines or non-parametric tests) will be 
conducted in future evaluations.

\subsection{Comparison with existing literature}
    The table~\ref{tab:comparison} presents the contribution of BC-TEAEM compared 
    to existing literature in the domain of blockchain-support recommendation 
    guided by user preferences.

    \begin{table}[ht]
    \centering
    \caption{Comparison of BC-TEAEM with Prior Work}
    \label{tab:comparison}
    \begin{tabular}{p{1.5cm}p{1.3cm}p{2cm}p{2.3cm}}
    \hline
    \textbf{Dimension} & \textbf{Prior Work} & \textbf{Limitation} & \textbf{BC-TEAEM} \\
    \hline
    Ontology & \cite{six2022ontology}\cite{xu2018pattern}\cite{xu2025pattern} & No link to soft goals & Extends ontology with soft goal relations \\
    \hline
    User Preferences & \cite{summa2023}\cite{bahar2023dynamic} & Not modeled explicitly & Captures and weights user-defined soft goals \\
    \hline
    Decision Support & \cite{six2022},\cite{Lui2022} & TOPSIS on static data & TOPSIS + dynamic soft goal influence matrix \\
    \hline
    Trade-offs & \cite{janovic} & Not modeled or implicit & Explicit inter-pattern and soft goal trade-offs \\
    \hline
    Tool Support & \cite{six2022ontology} & Conceptual or limited & Interactive tool with dynamic recommendations \\
    \hline
    \end{tabular}
    \end{table}

To complement our response to RQ$_2$ and RQ$_3$, our findings align with prior
work highlighting the importance of managing trade-offs in blockchain 
ecosystems~\cite{janovic} and the utility of soft goal modeling for capturing 
interdependencies~\cite{summa2023}\cite{hu2015semantic}. Unlike earlier MCDM 
approaches~\cite{bahar2023dynamic}\cite{Lui2022}, BC-TEAEM leverages 
ontology-based reasoning to improve understanding of inter-pattern and 
inter–soft goal relationships, thereby extending existing contributions with 
a domain-specific, tool-supported framework.


\section{Discussion and Threat to validity}\label{discussion}
\subsection{Discussion}
This section discusses the contributions of BC-TEAEM in relation to our 
research questions, with an emphasis on the extensibility of the artifact,
the blockchain pattern, and the underlying soft goal modeling approach.

\vspace{0.1cm}
\textit{Scope and Generalizability:} One of the key strengths of BC-TEAEM lies in its modular and domain-independent architecture. 
Although the current implementation is tailored to blockchain-based systems, the overall framework based on soft goal preferences, 
trade-off reasoning, and pattern recommendations is not restricted to any specific domain.
Soft goals can be redefined or extended, and domain-specific patterns can be integrated without altering the core decision-support logic.

\vspace{0.1cm}
While our case study centers on a pharmaceutical supply chain, the design 
of BC-TEAEM is not domain-specific. It can serve as reusable and 
extensible foundation for other  domains where trust-related issues are 
prominent, such as healthcare, supply chains, and finance.
By replacing or complementing the existing set of soft goals and patterns,
practitioners can apply the same reasoning process and recommendation 
workflow in new contexts. The ontology-driven structure also facilitates 
this extensibility by allowing the structured import of new concepts and 
relationships.

\subsection{Threats to validity:}
Despite its strengths, our approach has certain limitations that must be 
acknowledged. 

\vspace{0.1cm}
\noindent\textit{Ontology Coverage:} To ensure the ontology remains relevant 
and comprehensive, a systematic review of emerging blockchain design 
patterns should be performed. This would allow the continuous integration 
of the state-of-the-art patterns into the ontology, thus maintaining its 
accuracy and applicability in evolving domains.

\vspace{0.1cm}
\noindent\textit{Evaluation (Sample Size and Weights of Knowledge Base):}
The number of participants in the evaluation is limited, which may affect 
the generalizability of the results. With only a small panel of experts, 
the findings may not fully represent the broader research community or all
relevant domains. Also, the structure and relative importance assigned to 
elements in the knowledge base may introduce bias. If certain soft goals 
or patterns were over- or underrepresented, it could impact the outcomes 
and limit the neutrality of the reasoning process.

\vspace{0.1cm}
\noindent\textit{Interpretability and Decision Support:}
\noindent\textit{Interpretability and Decision Support:} While the system 
provides structured recommendations, non-expert users may struggle with 
complex trade-offs. Future work should integrate visual analytics and 
explainable decision-support features to improve transparency and clarify 
how patterns influence soft goals.


\vspace{0.1cm}
\noindent\textit{Bias in soft goal weighting:} The quality of recommendations 
depends on accurate and consistent weights from domain experts. Misjudgments 
or limited knowledge may cause divergence from real requirements. Future work 
could use structured elicitation techniques (e.g., pairwise comparisons, AHP)
to improve consistency and reduce subjectivity.


\section{Conclusion}\label{conclusion}
This paper introduced BC-TEAEM, a decision-support framework designed to assist system architects in selecting blockchain design patterns based on soft goals. 
Our proposal integrates three main components: an ontology that models the relationships between blockchain patterns and soft goals, 
a knowledge-based reasoning mechanism that captures interdependencies and trade-offs, 
and a MCDM method to prioritize alternatives based to stakeholder preferences.

\vspace{0.1cm}
By aligning business-level goals with technical design options, 
BC-TEAEM supports traceable and goal-oriented architectural decisions. 
The tool facilitates interaction between domain and technical experts, 
allowing iterative refinement of preferences and providing justifiable 
pattern recommendations.

\vspace{0.1cm}
An empirical evaluation involving  combining expert-based scenarios (expert interviews)
and user feedback highlights the relevance and usability of our approach.
Participants appreciated the tool’s ability to make trade-offs explicit and
 structure complex architectural choices.

\vspace{0.1cm}
Compared to existing approaches, BC-TEAEM offers novel contributions by
linking patterns to soft goals with directional influences, modeling user 
preferences, and operationalizing trade-off analysis within an interactive tool.
Although our evaluation is grounded in a supply chain case study, the 
framework itself is domain-independent. Future work will extend the ontology 
beyond supply chains, validating the framework in domains such as healthcare,
finance, and inter-organizational collaborations to demonstrate its 
generalizability. Furthermore, the reasoning engine will be enhanced with 
explainable AI features and assessed through large-scale industrial case 
studies.



\begin{thebibliography}{00}
\bibitem{bahar2023dynamic} Bahar, M. N., Ries, M., \& Košťál, K. (2023, October). A dynamic decision support system for the best-fitting blockchain platform selection. In 2023 Fifth International Conference on Blockchain Computing and Applications (BCCA) (pp. 412-419). IEEE.
\bibitem{Besancon2022} Besançon, L., Da Silva, C. F., Ghodous, P., \& Gelas, J.-P. (2022). A blockchain ontology for DApps development. \textit{IEEE Access}, 10, 49905--49933. doi:10.1109/ACCESS.2022.3173313
\bibitem{guizzardi2015towards} Guizzardi, G., Wagner, G., Almeida, J. P. A., \& Guizzardi, R. S. S. (2015). Towards ontological foundations for conceptual modeling: The unified foundational ontology (UFO) story. \textit{Applied Ontology}, 10(3-4), 259--271.
\bibitem{henry2024} Henry, T., \& Tucci-Piergiovanni, S. (2024, September). Secure proof verification blockchain patterns. In International Conference on Business Process Management (pp. 71-88). Cham: Springer Nature Switzerland.
\bibitem{hu2015semantic} Hu, H., Ma, Q., Zhang, T., Tan, Y., Xiang, H., Fu, C., \& Feng, Y. (2015). Semantic modelling and automated reasoning of non-functional requirement conflicts in the context of soft goal interdependencies. IET Software, 9(6), 145--156. Wiley Online Library.
\bibitem{janovic} Jovanovic, M., Kostić, N., Sebastian, I. M., \& Sedej, T. (2022). Managing a blockchain-based platform ecosystem for industry-wide adoption: The case of TradeLens. Technological Forecasting and Social Change, 184, 121981.
\bibitem{kambilo2023} Kambilo, E. K., Rychkova, I., Herbaut, N., \& Souveyet, C. (2023, May). Addressing trust issues in supply-chain management systems through blockchain software patterns. In International Conference on Research Challenges in Information Science (pp. 275-290). Cham: Springer Nature Switzerland.
\bibitem{Lui2022}Liu, J., Shen, X., Xie, M., \& Zhang, Q. (2022, November). Research on sharding strategy of blockchain based on TOPSIS. In International Conference on Smart Computing and Communication (pp. 247-257). Cham: Springer Nature Switzerland.
\bibitem{tam2015} Marangunić, N., \& Granić, A. (2015).  Technology acceptance model: A literature review from 1986 to 2013.  \textit{Universal Access in the Information Society}, 14, 81--95.  https://doi.org/10.1007/s10209-014-0348-1
\bibitem{papathanasiou2018topsis} Papathanasiou, J., \& Ploskas, N. (2018). \textit{TOPSIS}. Springer.
\bibitem{rajasek} Rajasekar, V., Sondhi, S., Saad, S., \& Mohammed, S. (2020, July). Emerging Design Patterns for Blockchain Applications. In ICSOFT (pp. 242-249).
\bibitem{six2022} Six, N., Herbaut, N., \& Salinesi, C. (2022). Blockchain software patterns for the design of decentralized applications: A systematic literature review. \textit{Blockchain: Research and Applications}, 3(2), 100061. Elsevier.
\bibitem{six2022ontology} Six, N., Correa-Restrepo, C., Herbaut, N., \& Salinesi, C. (2022).  An ontology for software patterns: Application to blockchain-based software development. In \textit{International Conference on Enterprise Design, Operations, and Computing} (pp. 284--299). Springer.
\bibitem{summa2023} Soundararajan, R., \& Shenbagaraman, V. M. (2023). Unlocking the Potential of Blockchain Through Multi-Criteria Decision Making in Platform Selection. International Journal of Professional Business Review: Int. J. Prof. Bus. Rev., 8(4), 49
\bibitem{sumesh2022} Sumesh, S., \& Krishna, A. (2022). Challenges and review of goal-oriented requirements engineering based competitive non-functional requirements analysis. \textit{Multiagent and Grid Systems}, 18(2), 171--191. SAGE Publications Sage UK: London, England.
\bibitem{sus2024} Vlachogianni, P., \& Tselios, N. (2022). Perceived usability evaluation of educational technology using the System Usability Scale (SUS): A systematic review. \textit{Journal of Research on Technology in Education}, 54(3), 392--409. https://doi.org/10.1080/15391523.2021.1929032
\bibitem{wieringa2014design} Wieringa, R. (2014). ``Design Science Methodology for Information Systems and Software Engineering``'. Springer.
\bibitem{xu2018pattern} Xu, X., Pautasso, C., Zhu, L., Lu, Q., \& Weber, I. (2018).  A pattern collection for blockchain-based applications. In Proceedings of the 23rd European Conference on Pattern Languages of Programs (pp. 1--20).
\bibitem{xu2025pattern} Xu, X., Pautasso, C., Lo, S. K., Zhu, L., Lu, Q., \& Weber, I. (2025). An Extended Pattern Collection for Blockchain-Based Applications. In Transactions on Pattern Languages of Programming V (pp. 67-117). Berlin, Heidelberg: Springer Berlin Heidelberg.

\end{thebibliography}
\end{document}